\documentclass{PoS}

\usepackage{aas_macros}
\usepackage{lineno}

\title{Highlights from the High Altitude Water Cherenkov Observatory}

\ShortTitle{HAWC Highlights}

\author{\speaker{John Pretz}$^a$ for the HAWC Collaboration$^b$ \\
        \llap{$^a$}Department of Physics, Pennsylvania State University, State College, PA, USA \\
        \llap{$^b$}For a complete author list, see \href{http://www.hawc-observatory.org/collaboration/icrc2015.php}{www.hawc-observatory.org/collaboration/icrc2015.php} \\
        Email: \email{pretz@psu.edu}}

\abstract{The High Altitude Water Cherenkov (HAWC) Gamma-Ray Observatory 
was completed this year at a 4100-meter site on the flank of the Sierra 
Negra volcano in Mexico. HAWC is a water Cherenkov ground array with the 
capability to distinguish 100 GeV - 100 TeV gamma rays from the hadronic 
cosmic-ray background. HAWC is uniquely suited to study extremely high 
energy cosmic-ray sources, search for regions of extended gamma-ray emission, 
and to identify transient gamma-ray phenomena. HAWC will play a key role in 
triggering multi-wavelength and multi-messenger studies of active galaxies, 
gamma-ray bursts, supernova remnants and pulsar wind nebulae. Observation 
of TeV photons also provide unique tests for a number of fundamental 
physics phenomena including dark matter annihilation and primordial black 
hole evaporation. 
Operation began mid-2013 with the partially-completed 
detector. Multi-TeV emission from the Galactic Plane is clearly seen in the 
first year of operation, confirming a number of known TeV sources, and a 
number of AGN have been observed. We discuss the science of 
HAWC, summarize the status of the experiment, and highlight first results 
from analysis of the data.}

\FullConference{The 34th International Cosmic Ray Conference,\\
		30 July- 6 August, 2015\\
		The Hague, The Netherlands}

\begin{document}

\section{Introduction}

The High Altitude Water Cherenkov Observatory (HAWC) was completed in early 
2015 and is currently surveying the Northern sky for gamma-ray sources between 
100 GeV and 100 TeV. HAWC is distinctive for its sensitivity to extreme
high-energy photons, wide 
field-of-view and ability to operate continuously. In one year, 
HAWC will survey the entire overhead sky above $\sim$ 5 TeV with the 
sensitivity \cite{hawcsteadysourcesensitivity} of a 50-hour 
pointed observation by the current
generation of Imaging Atmospheric Cherenkov Telescopes.

\begin{figure}
\includegraphics[width=0.95\linewidth]{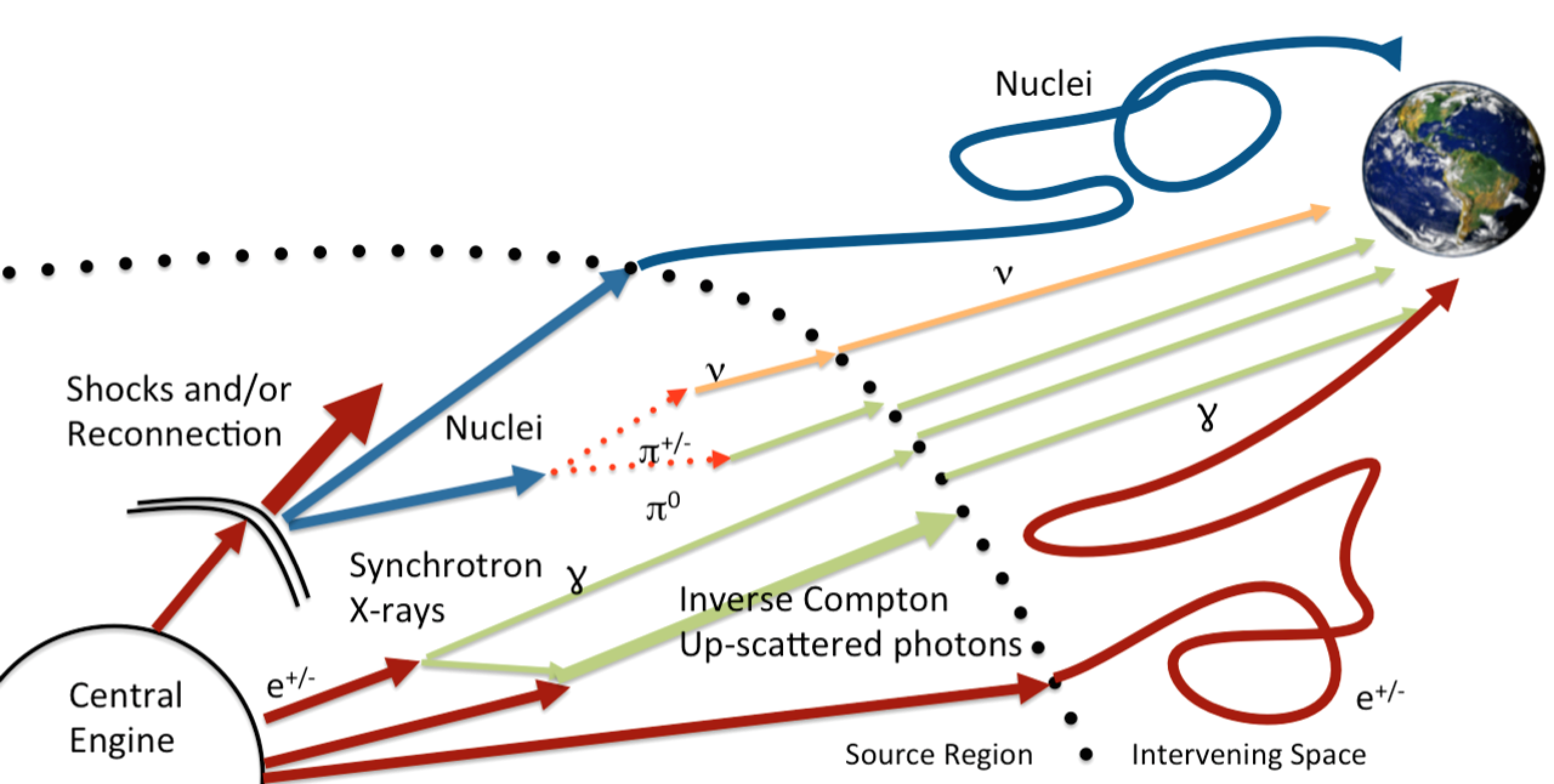}
\caption{
A schematic overview of the standard picture of a cosmic accelerator.
High-energy electrons and hadronic cosmic-rays are accelerated in a
source. The messengers received at Earth can be either these 
charge particles directly propagating from the source, or neutral
secondaries (photons or neutrinos) produced in pion decay or inverse
Compton scattering within the source. 
}
\label{overview}
\end{figure}

The central objective of HAWC is to study particle acceleration in 
high-energy sources and subsequent propagation of these particles to Earth. 
Figure \ref{overview} shows a schematic of the standard picture of particle
acceleration in sources. The chief complication in modeling these sources
is whether hadronic cosmic rays or electrons and positrons are responsible
for the observed gamma rays. 
HAWC's measurements can illuminate the 
nature of these accelerators by characterizing the highest energy these
sources can achieve, the long-duration
time structure of the high-energy emission, triggering follow-up by 
partner observatories, 
and by measuring the extended emission as the high-energy particles stream
from their sources.

The HAWC science goals include:

\begin{itemize}

\item Distinguishing gamma-ray emission caused by electron acceleration or 
hadronic cosmic-ray acceleration is a persistent challenge in modeling 
accelerators.  In order to build a complete picture of a source, it is critical 
to know the highest energy particles a source can accelerate. HAWC has 
unprecedented sensitivity above 10 TeV and will be used to measure gamma-ray 
spectra of Galactic sources up to 100 TeV. 

\item The rapid variability of some sources points to acceleration in regions 
that are too small to be spatially resolved by gamma-ray instruments. With 
HAWC, we continuously monitor the sky for bright $\sim$TeV outbursts from 
Active Galaxies, Gamma-Ray Bursts or Galactic transients, enabling 
long-duration TeV light curves and
multi-wavelength followup and study.

\item A number of large gamma-ray structures have been seen in the Galaxy, 
including extended emission in the Galactic plane, most notably in the Cygnus
arm of the Galaxy, in the region around the Geminga pulsar, and the
two extended lobes of gamma-rays (the Fermi Bubbles). With HAWC, 
we have an unbiased survey of the Northern sky, 
and a full accounting of all the 
TeV photons in the sky. 

\item HAWC will search for a number of signatures of new physics, including 
Dark Matter annihilation, Lorentz Invariance Violation and Primordial Black 
Hole evaporation. 

\item Hadronic cosmic-rays at Earth constitute the chief background to 
gamma-ray observation. Nevertheless, HAWC data can be used to study the 
population of 100 GeV - 100 TeV hadronic cosmic rays directly, and can use 
the cosmic rays as a probe of solar physics, through the sun's modulation
of the cosmic rays at Earth.
Spatial anisotropies of cosmic-rays are not well understood and may probe 
local magnetic field structure.

\end{itemize}



Section \ref{detectorsection} discusses the state of the HAWC detector.
HAWC scientific efforts are divided into four categories described
in the following sections:
In Section \ref{galacticsection}, we present results on gamma-ray 
studies of
objects in the Galaxy. Section \ref{extragalacticsection} discusses
gamma-ray observations of objects outside our galaxy.
Section \ref{fundamentalsection} presents our efforts to constrain
fundamental physics using astrophysical gamma-ray observations.
Section \ref{crsection} discusses our efforts to study the cosmic-rays
directly, including the impact of the sun on the cosmic-ray population.

\section{HAWC Instrument}
\label{detectorsection}

\begin{figure}
\includegraphics[width=0.95\linewidth]{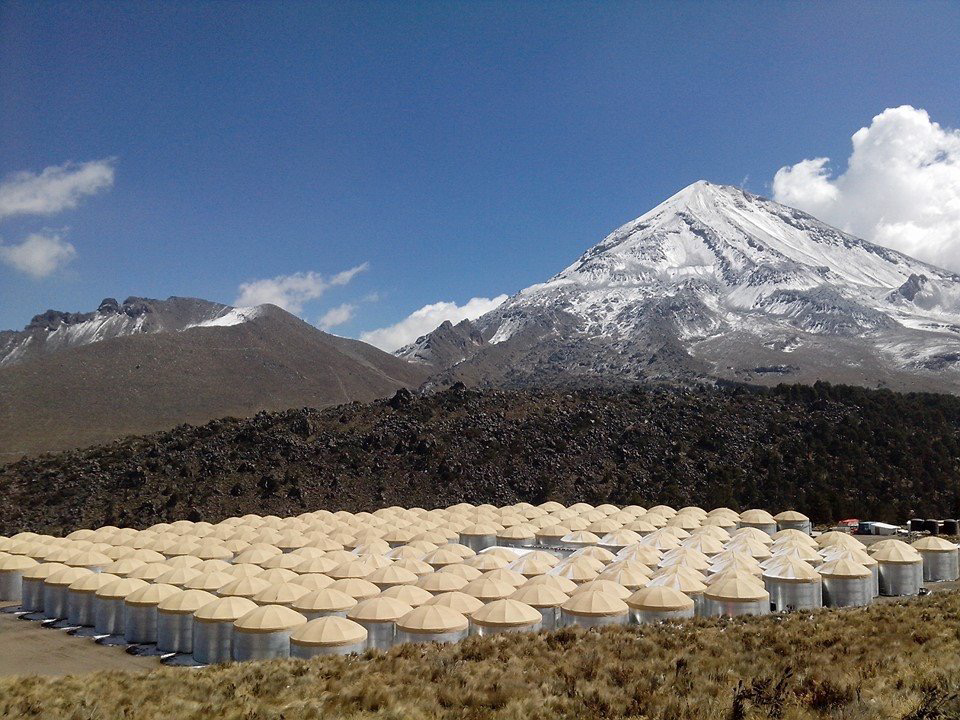}
\caption{
The High Altitude Water Cherenkov Observatory at its home near
Pico de Orizaba in Mexico. The 300 WCDs are large and 
close-packed giving the instrument a footprint of 20,000 m$^2$.
}
\label{hawcbaby}
\end{figure}

Figure \ref{hawcbaby} shows the completed HAWC detector at the
4100 meter altitude site near Pico de Orizaba in Mexico. 
The HAWC instrument consists of an array of 300 water Cherenkov
detectors (WCDs). Each WCD comprises a steel tank 7.3 meters in diameter and
4.5 meter height, a plastic bladder to contain 188,000 liters of purified water,
and four PMTs: three 8-inch Hamamatsu R5912 PMTs re-used from Milagro
and one 10-inch R7081-MOD high-quantum efficiency PMT. The WCDs are 
deployed in a close-packed array over an area of approximately 20,000 m$^2$.
Cables connect the PMTs to a central counting house, deliver high
voltage to the PMTs, and carry signals to the acquisition electronics.
 A calibration system
\cite{calibrationproceedings}
allows the delivery of light via optical cable to the WCDs in order to measure
the length of the cables and the relationship between the amount of light delivered
and the measured signal in PMTs.

The PMTs detect Cherenkov light from energetic particles passing through the
WCDs during cosmic-ray and gamma-ray air showers.
 The air shower reconstruction takes place in three steps: First,
the air shower core, the dense concentration of energetic particles
directly
along the primary particle's trajectory, is found with a fit
to the energy density recorded in PMTs on the ground. Second, 
the air shower direction is determined.
Air showers propagate near the speed of light, so
the
energetic particles arrive nearly in a plane, with particles far from
the core arriving somewhat delayed.
The PMT hit times are fit to this hypothesis to determine
the shower direction.
In a final step, a
number of event parameters are computed to determine if the event is
likely
to be a photon- or hadron-initiated particle cascade.

\begin{figure}
\includegraphics[width=0.45\linewidth]{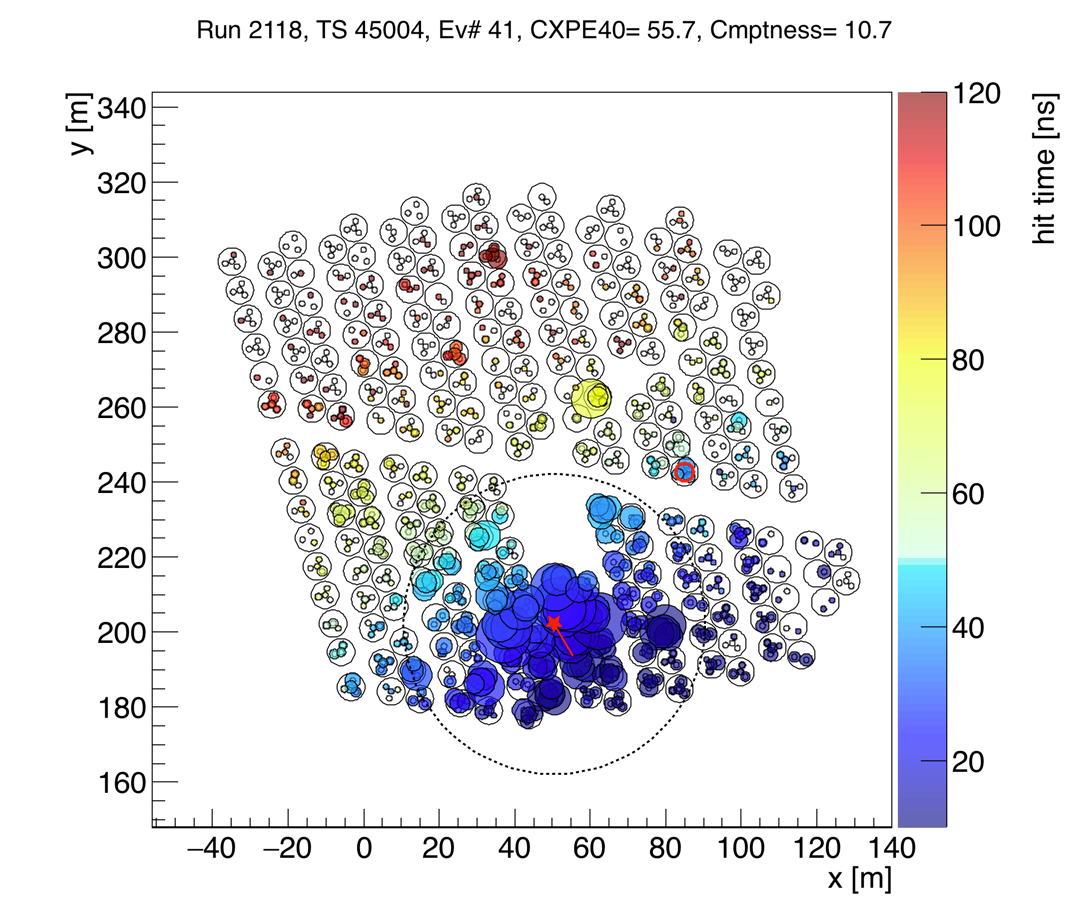}
\includegraphics[width=0.45\linewidth]{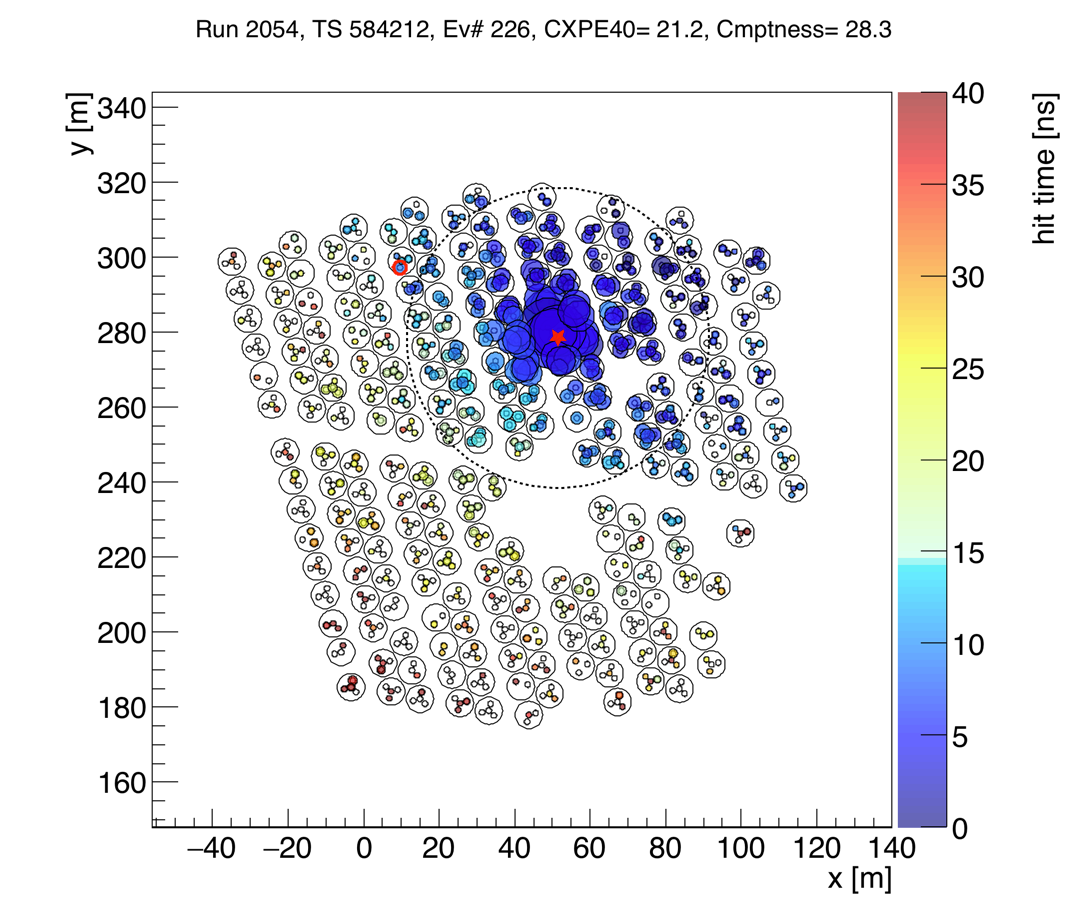}
\caption{
Two events recorded in HAWC. Each point
represents a PMT in HAWC and the WCD tank boundaries are shown. The
color indicates the time each PMT saw light and the the size indicates
how much light was seen. In each case, the shower core is readily 
apparent as the dense region of hard-hit PMTs.  The left
event is a hadronic cosmic-ray event, with a number of regions of
isolated energy deposition outside the shower core. The right figure
came from the direction of the Crab nebula and is likely
to be a true $\sim$ 10 TeV photon. Note the absence of energy deposition
far from the core for the photon, denoted by the 40-meter circle drawn 
around the core.
}
\label{events}
\end{figure}

Figure \ref{events} shows
two events recorded in HAWC. In both cases, the air shower core is
readily
discernible as well as the planar time distribution. The
hadron-initiated
event in the left panel has a number of isolated regions of
high-energy
particles well outside the core region. The gamma-ray event in the
right
panel is much more smooth, with no large charge deposition outside the 
shower core region.  These characteristics form the basis for the
current
photon/hadron discrimination in HAWC \cite{andysproceedings}, and
additional
advanced separation techniques are being investigated
\cite{zigproceedings,tomasproceedings}.

The chief limitation of HAWC's sensitivity at energies above about 10
TeV is the collection area of the instrument. To first order, air
shower particles arrive in a plane defined by the speed of light, but
particles far from the shower axis arrive 
delayed.  The effect is substantial, up to tens of ns 100 meters from 
the air shower core.
The accurate reconstruction of showers
requires correction of this effect, and the correction is done
imperfectly if the air shower core is well off the main instrumented
area.
Off-instrument events are usable. The
separation of photons and hadrons can be done, but without 
a reliable core localization, the direction cannot be accurately determined.
In order to recover these events, a new sparse outrigger array is planned
\cite{orproceedings}. Such an array will, at modest additional cost,
greatly
enhance the high-energy sensitivity of HAWC.

\begin{figure}
\includegraphics[width=0.95\linewidth]{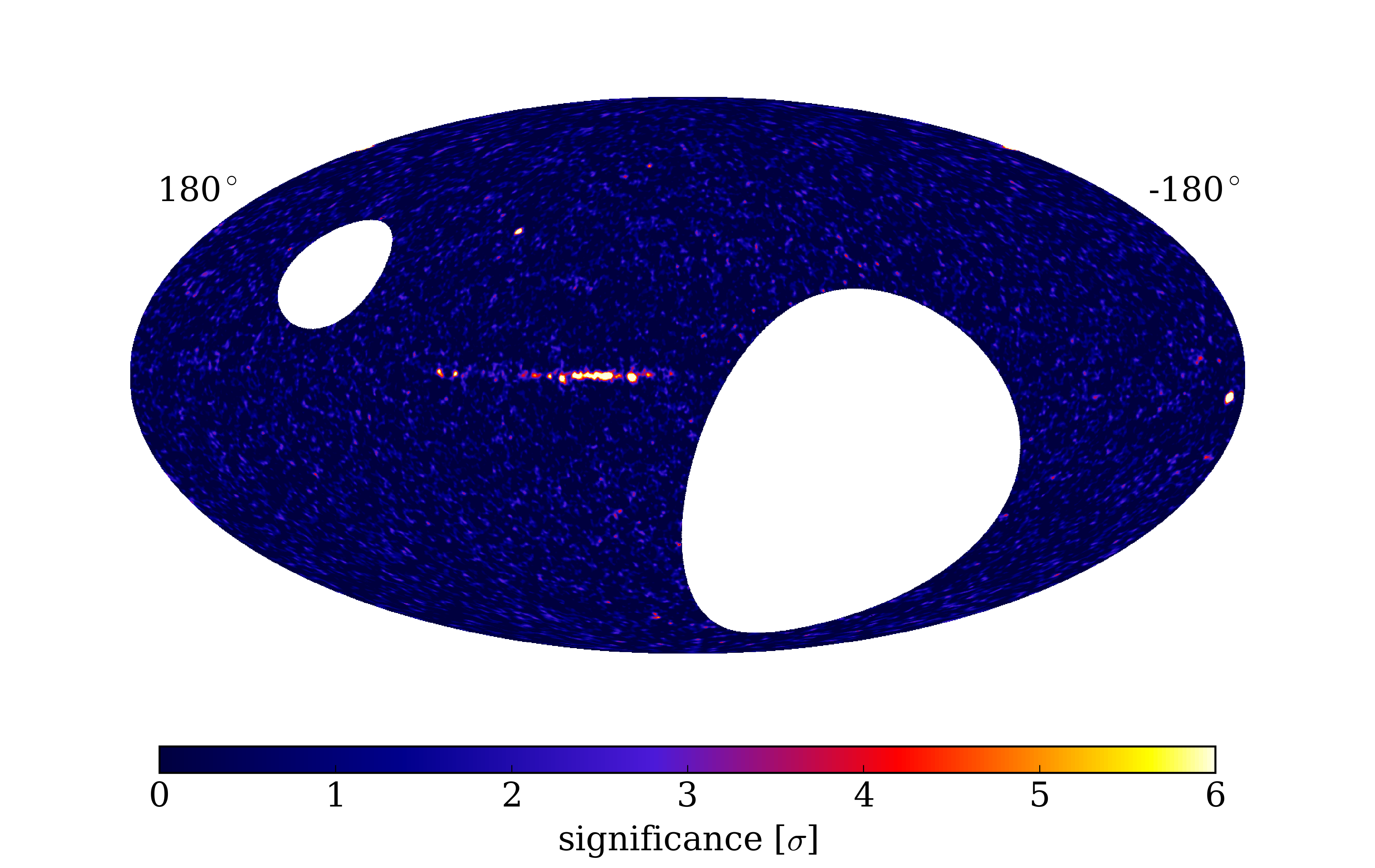}
\caption{
The all-sky skymap from 150 days of the nearly-completed HAWC instrument.
The sky is shown in Galactic coordinates. TeV emission from a number
of source regions in the Galactic plane are evident. 
The Crab is seen to the far right. The TeV blazars Markarian 
501 and 421
are detected at high significance. Markarian 501 is conspicuously located
above the plane of the galaxy. Markarian 421, in this projection, is 
right at the edge of the field, near the top.
}
\label{allsky}
\end{figure}

After air showers have been reconstructed, we make a map of the sky. 
Figure \ref{allsky} shows a significance map from the
first 150 days of the near-final HAWC
instrument in Galactic coordinates. Clear emission along the plane of
the galaxy is seen, along with emission from the Crab Nebula, and the 
TeV blazars Markarian 421 and 501. 

This paper presents data from two datasets:

\begin{itemize}

\item {\bf{HAWC-111}}: 
Prior to the completion of HAWC, the instrument was operated
in a partial configuration. HAWC-111 refers to the instrument between
August 2, 2013 and July 7, 2014, during which between 106 and 133
of the total 300
WCDs were operational, for a total livetime of 283 days.

\item {\bf{HAWC}}: HAWC began operating in its near-final 
configuration with 247 of its 300 WCDs live 
on November 26, 2014. We present data from 149 live days.

\end{itemize}

The integral sensitivity of the full HAWC sample is almost double the
sensitivity of the HAWC-111 sample, but the understanding of the HAWC-111,
and the evaluation of its systematics, is more mature. Some analyses
are presented from the HAWC-111 epoch and some from the full HAWC 
operation.

\section{Galactic Science}
\label{galacticsection}

HAWC data clearly shows TeV emission along the Galactic plane including a number of sources and 
extended regions of emission. 

The Crab Nebula, the standard candle
in TeV astronomy, has been observed at high confidence \cite{crabproceedings}.
With the instrument in its near-final 
configuration, we are observing the Crab at more than 3$\sigma$ each day,
for a total significance of $\sim$38$\sigma$ in the first 150 days of
data.

\begin{figure}
\includegraphics[width=0.45\linewidth]{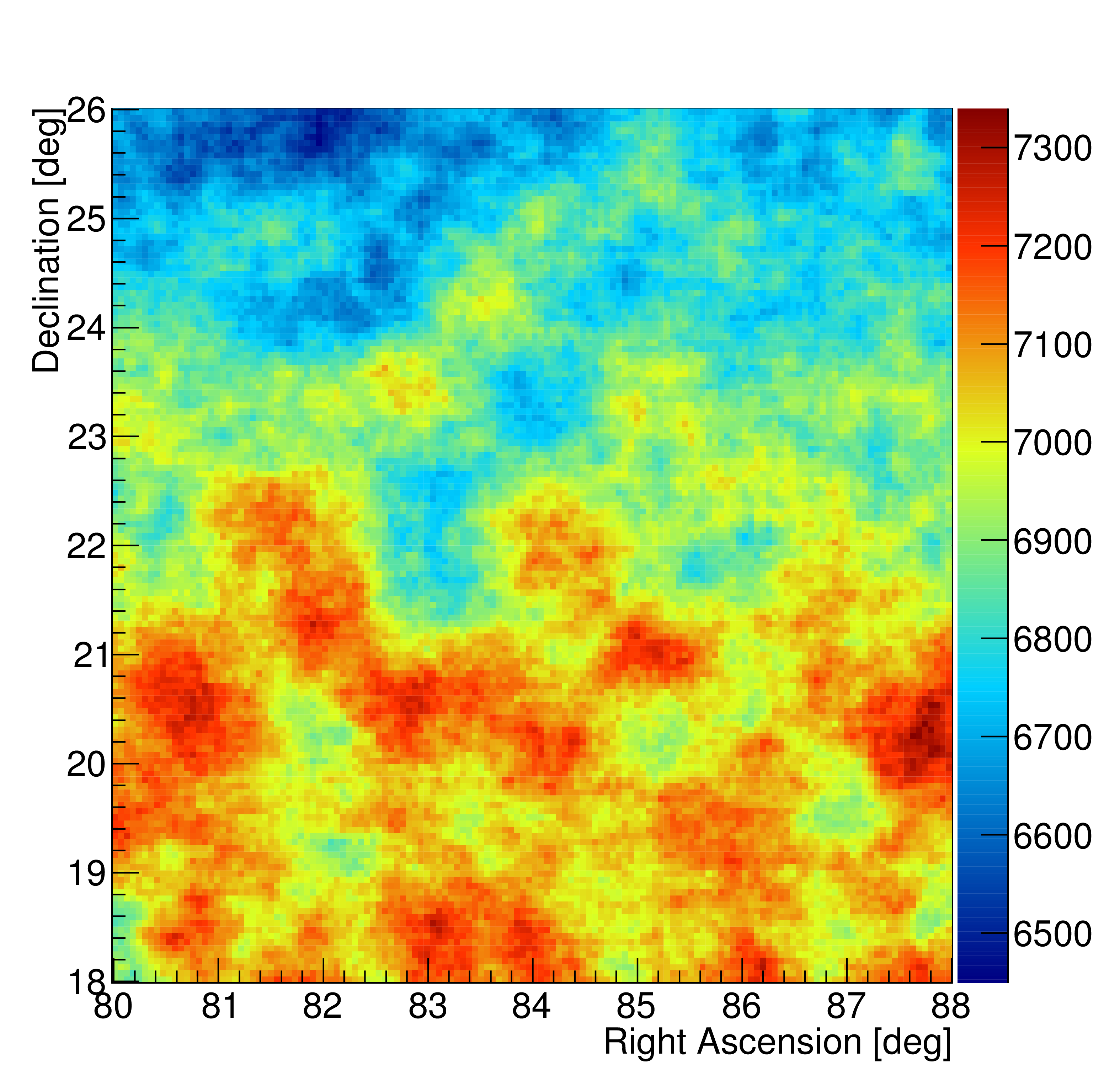}
\includegraphics[width=0.45\linewidth]{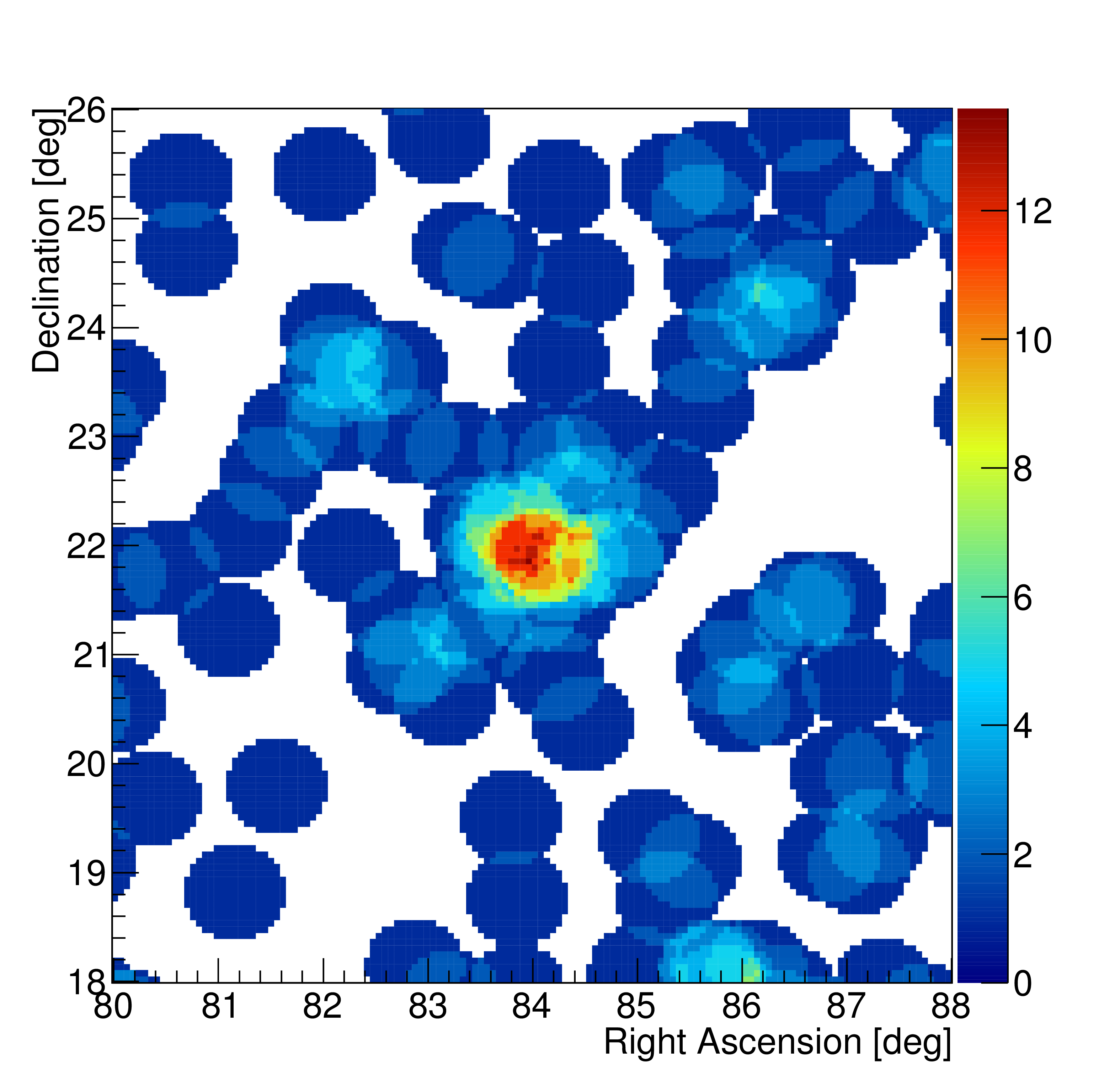}
\caption{
Skymap in equatorial coordinates in the vicinity of the Crab Nebula
in 105 days of data with HAWC-250 (shown
in the J2014/15 epoch) from \cite{isotropicproceedings}.
Only events with more than 85\% of the available PMTs are shown.
The color scale shows the number of events detected within a
0.45$^\circ$ circle around each point in the sky.
The left figure shows the region with no cuts and is completely
dominated by hadronic cosmic rays. The slight
drift in rate is due to the changing acceptance of the
detector, located at +19$^\circ$N latitude, for events further
from zenith. The figure on the right
shows the same skymap after a strong cut which removes all but 1 in $\sim$10$^4$
background events.
After cuts, at the location of the Crab,
we observe 12 events with an expected background of 0.81$\pm$ 0.05
in this sample.
}
\label{crabhighpurity}
\end{figure}

The Crab detection is particularly important because it allows us to
validate the modeling and performance of the instrument. In
particular, we are able to generate an extremely pure photon sample
and can study the instrument response to photons of known direction.
Figure
\ref{crabhighpurity} shows the field-of-view of the Crab for extremely
large events (the easiest to distinguish from hadronic
background). After
photon/hadron cuts, we have a signal to background ratio greater than
10. This is unprecedented for a wide-field ground array.
The strong hadron rejection in the vicinity of the Crab has been
used to place an upper limit on isotropic photon emission above 10 TeV.
\cite{isotropicproceedings}.
Currently, HAWC data is
within a factor of 2 of the design sensitivity and is improving with
our understanding of the instrument. 

\begin{figure}
\includegraphics[width=0.95\linewidth]{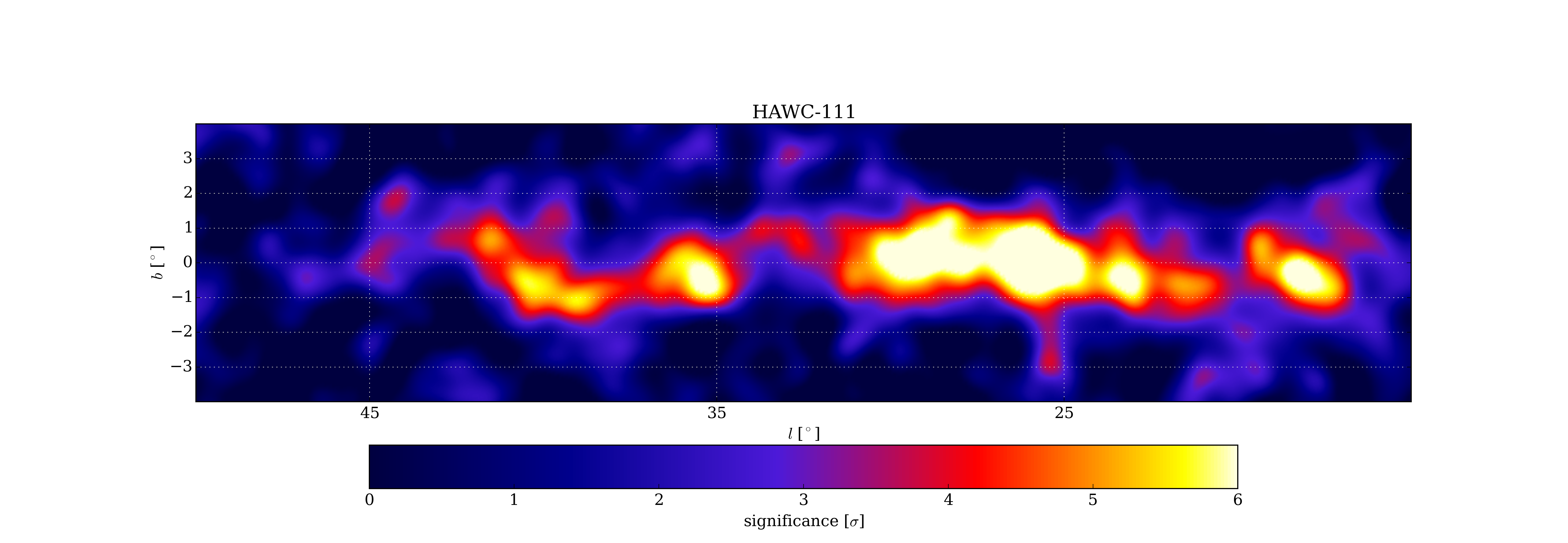}
\includegraphics[width=0.95\linewidth]{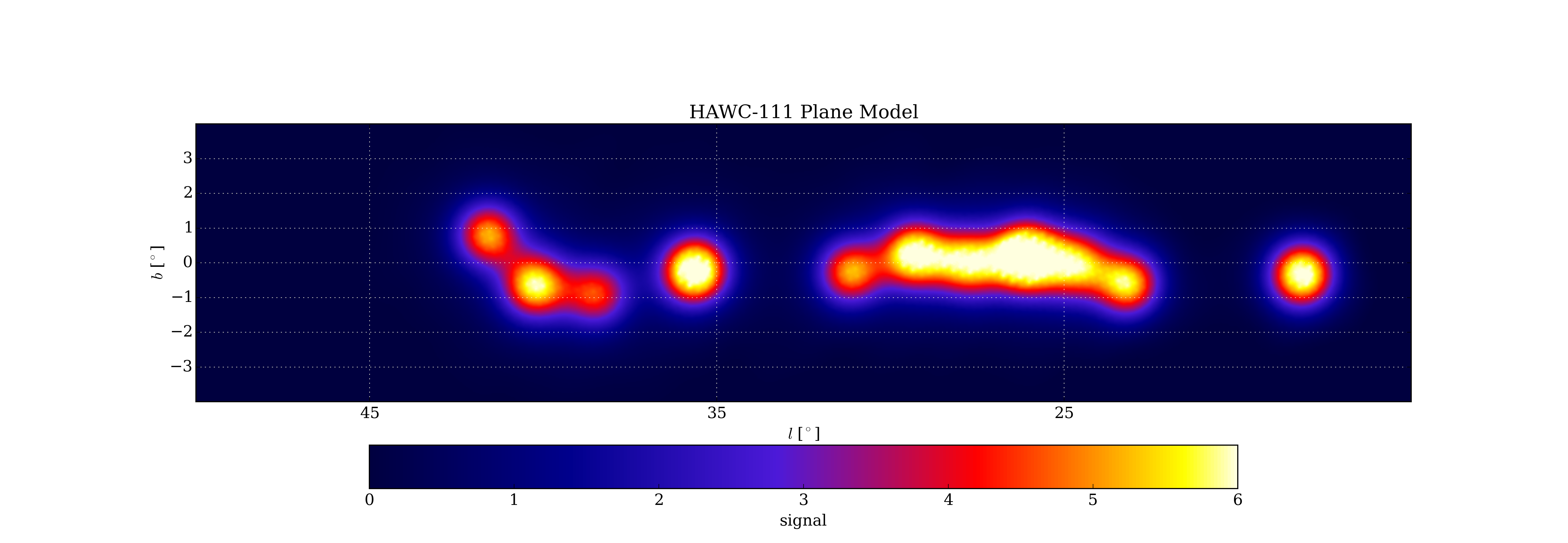}
\includegraphics[width=0.95\linewidth]{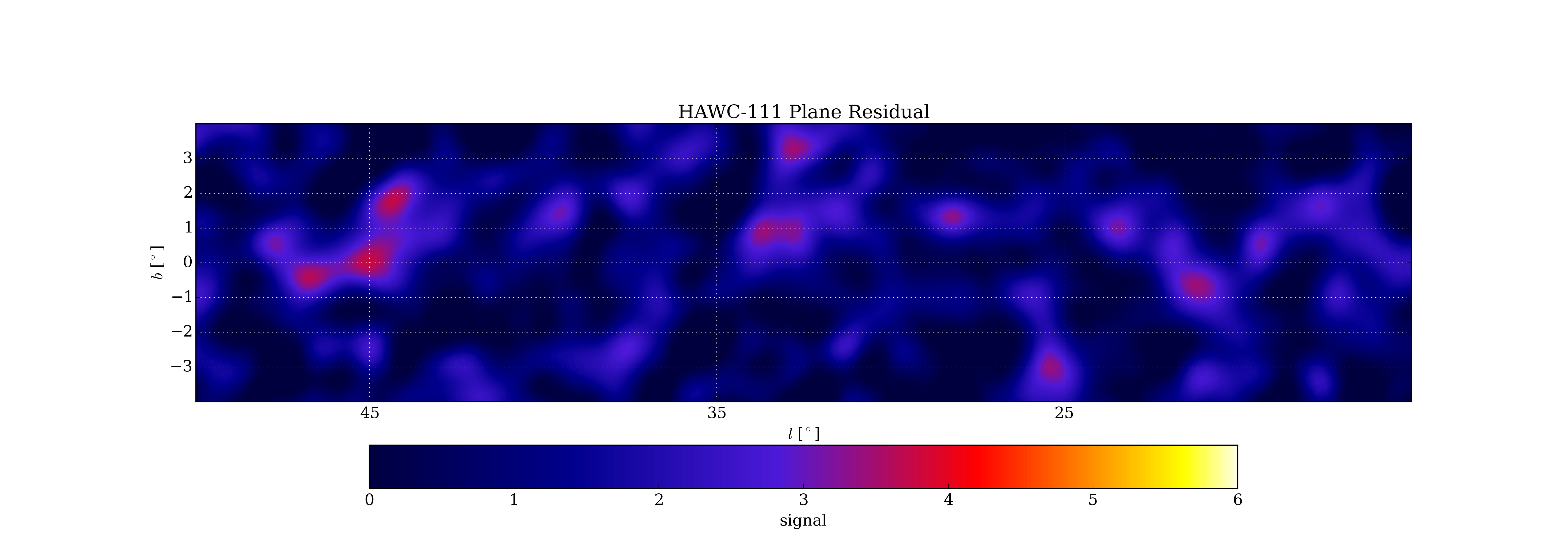}
\caption{
The Galactic plane survey in HAWC-111 
data \cite{planeproceedings,snrpwnproceedings}. The top figure shows the
significance map of the inner region of the Galactic plane. The
middle figure shows the 11-source model developed which explains 
the emission, resulting in 10 sources and source candidates. The bottom
figure shows the residual emission, showing that all significant
point sources have been successfully removed.
}
\label{plane}
\end{figure}

Figure \ref{plane} shows a region
of the Galactic plane visible to HAWC. The data is shown from
283 days of data in the HAWC-111 configuration.  An analysis has been
developed \cite{liffproceedings} which allows us to model the region
as an ensemble of point sources \cite{hawcbsl}. This results in the detection of 10
sources
and source candidates
\cite{snrpwnproceedings,planeproceedings}, most of which are
associated with known TeVCat \cite{tevcat} sources.  
Searches for Galactic 
transients \cite{binariesproceedings,pulsarproceedings} are in progress. 

In addition to the region of the inner Galactic plane, 
HAWC data also reveals a region of extended emission around the Geminga
pulsar \cite{gemingaproceedings}. Geminga is nearby (about 250 pc) and has
the potential to explain $\sim$GeV positron excess \cite{amspositronexcess} 
as direct production of
electrons and positrons \cite{gemingatheory}.
TeV emission around Geminga was first
observed with Milagro \cite{milagrofermibsl}, but the source's large
extent has defied identification by TeV IACTs. In HAWC data, the
source appears very large. More data is required to fit an extent
with reliable errors, 
but a $3^\circ$ round bin maximizes the
statistical significance of the detection at 6.3$\sigma$
pre-trials. 

\begin{figure}
\includegraphics[width=0.45\linewidth]{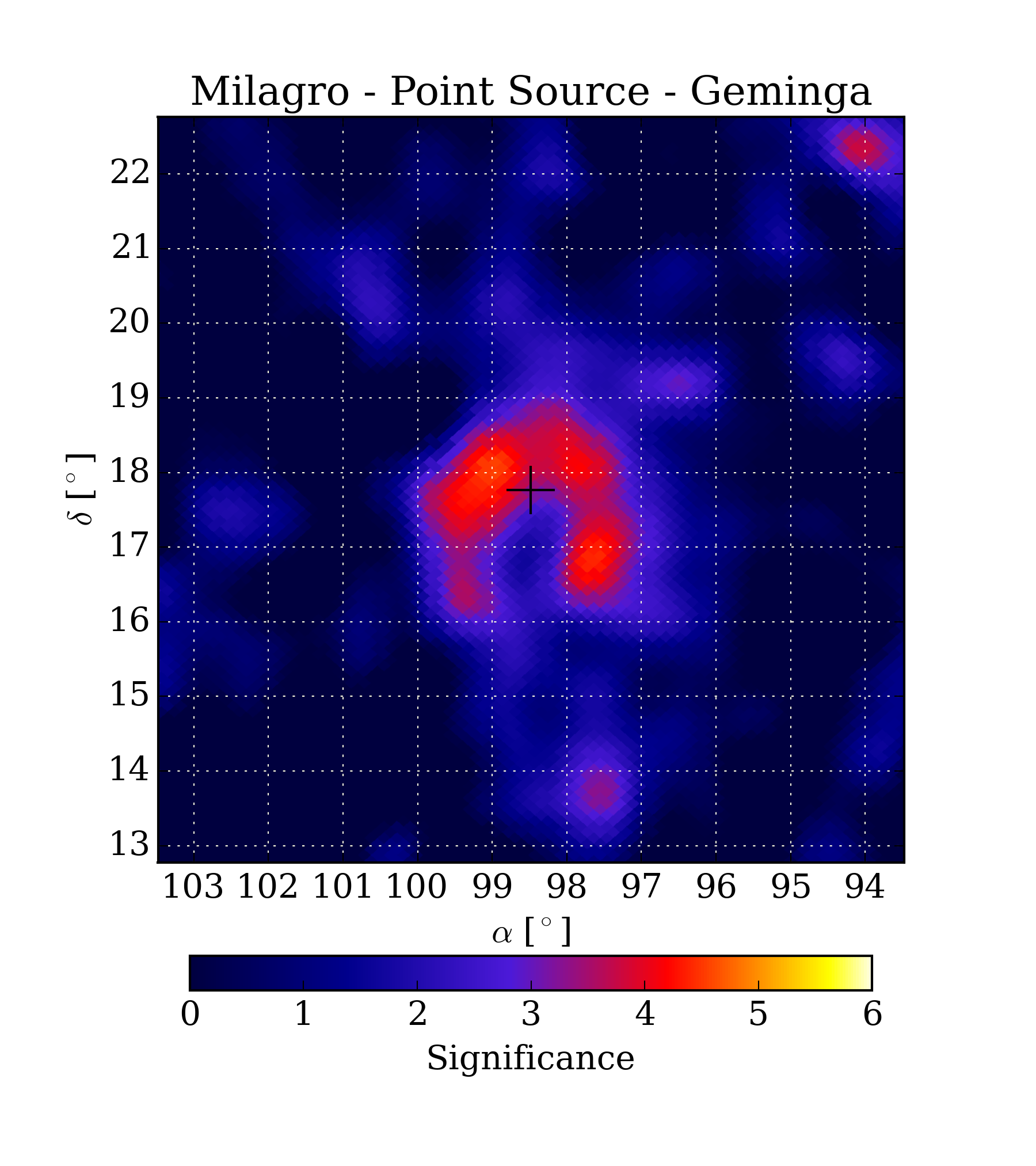}
\includegraphics[width=0.45\linewidth]{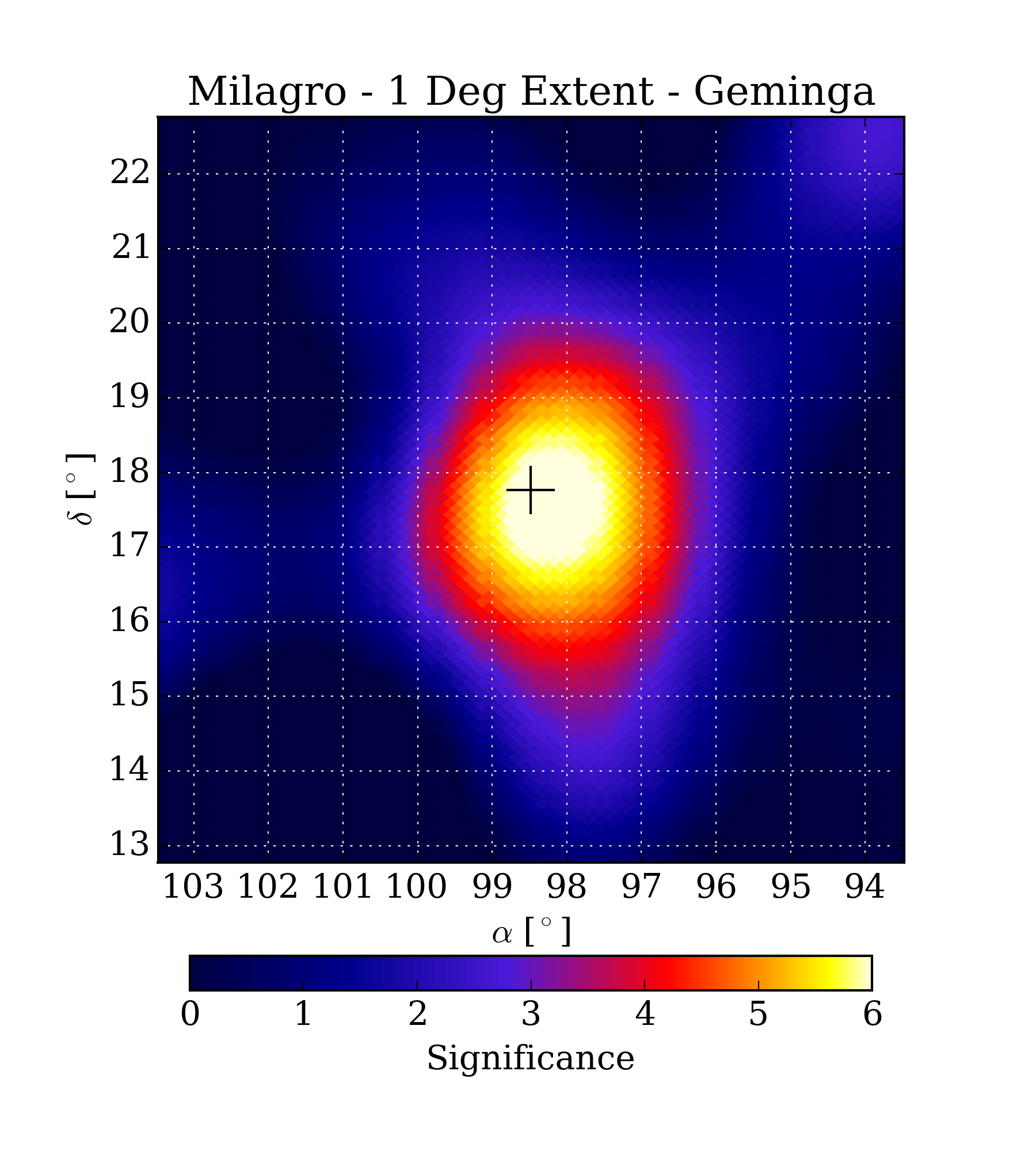} \\
\includegraphics[width=0.45\linewidth]{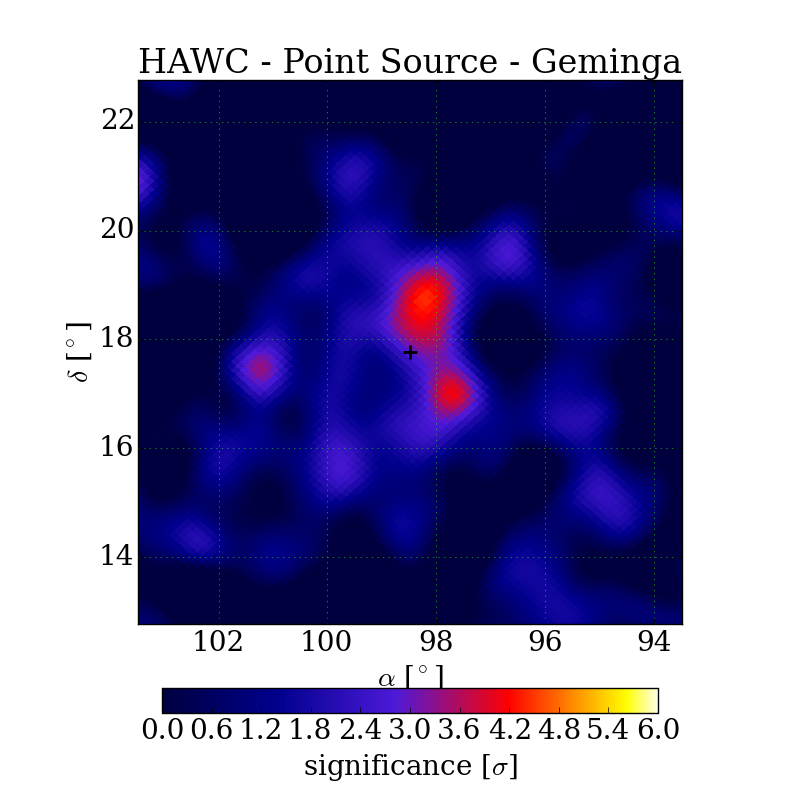}
\includegraphics[width=0.45\linewidth]{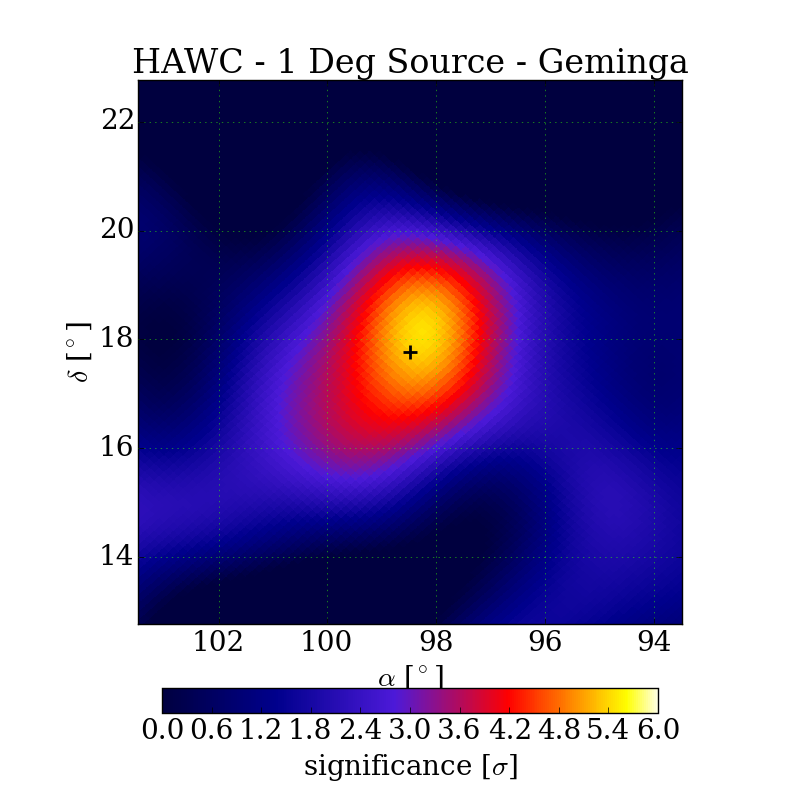}
\caption{
HAWC and Milagro observation of Geminga \cite{gemingaproceedings}. 
The region of TeV emission
around Geminga is large and extended. The top (bottom)
leftmost figure shows the 
Milagro (HAWC) statistical significance map, assuming the source
is a point source. The region is much more significant under the
assumption that the source is extended by 1 degree, shown on the right.
}
\label{geminga}
\end{figure}

\begin{figure}
\includegraphics[width=0.95\linewidth]{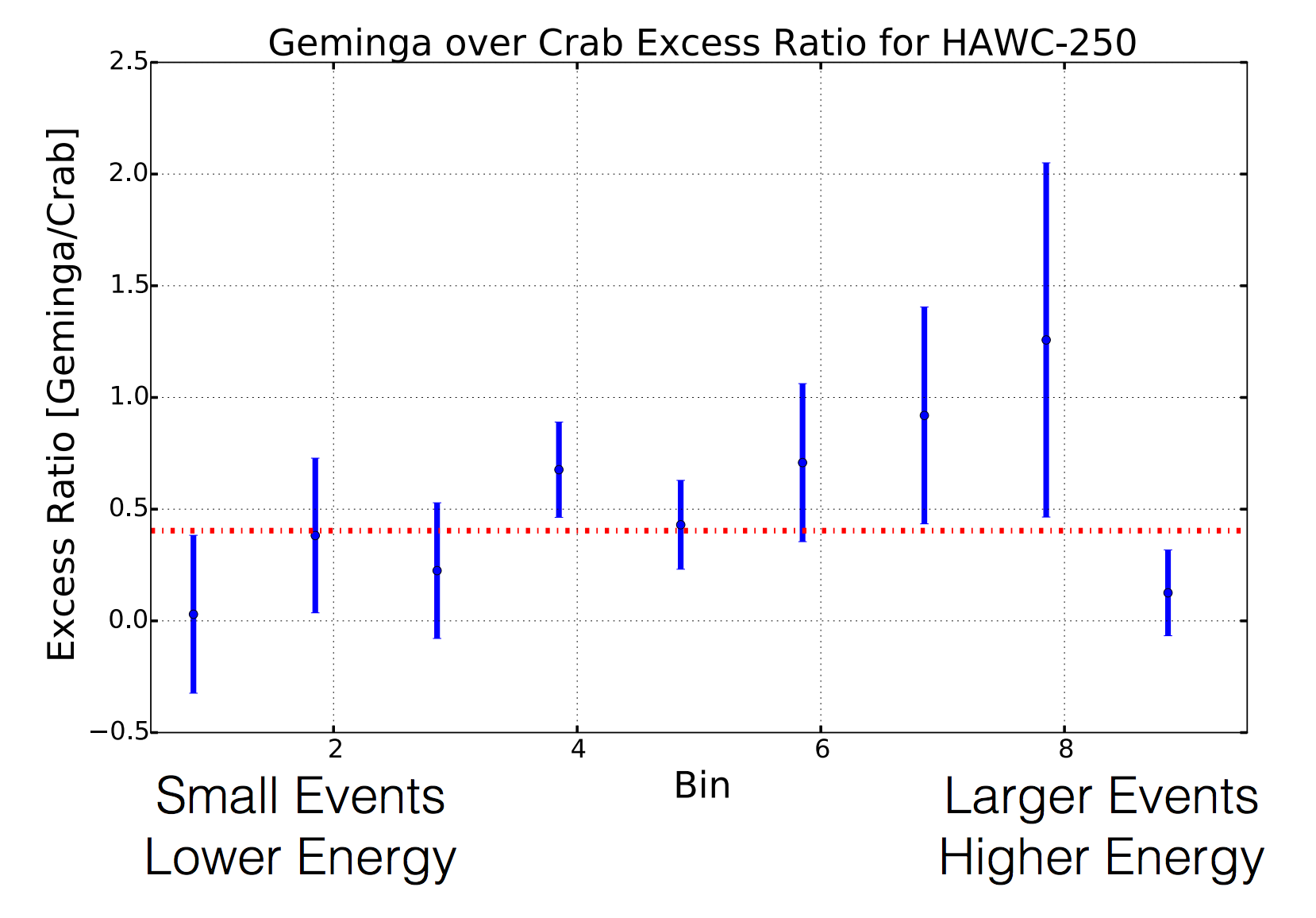}
\caption{
The relative gamma-ray excess from the Geminga and the Crab in 
data from the nearly-completed HAWC detector \cite{gemingaproceedings}. 
The excess is shown
as a function of "analysis bin" which is a proxy for the energy of 
gamma-ray events. The gentle upward trend suggests that the source
region may be harder than the Crab with a total surface brightness 
approximately
40\% of the Crab.
}
\label{gemingaexcess}
\end{figure}

The analysis of the excess around Geminga is in progress, but the
source, 
at a declination of $\sim$18$^\circ$, is near the Crab
Nebula. This proximity permits direct comparison to the Crab,
with the most relevant systematic issues canceling. Figure
\ref{gemingaexcess}
shows the number of excess events in a large 3$^\circ$ region around
Geminga
as a function of the size of the events in HAWC, relative to the
same analysis applied to the Crab. Larger events are
higher energy with ``Bin 0'' events just below 1 TeV and ``Bin 9''
events
at $\sim$10 TeV. The total number of photons from Geminga
is large, approximately 40\% of the total brightness of the Crab.
If the Geminga data were flat in this figure, we
would conclude it had the same spectral properties as the Crab. The
slight trend upward, while not significant, suggests a harder spectrum
than the Crab. It is worth noting a hard spectrum was predicted for
Geminga 
in \cite{gemingatheory}. A more complete treatment,
and more exposure
is necessary, though, to measure the spectrum accurately.

\begin{figure}
\includegraphics[width=0.95\linewidth]{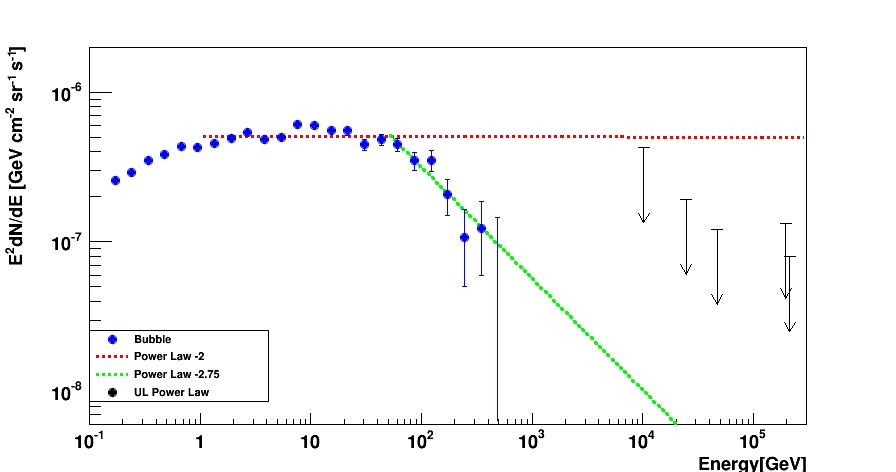}
\caption{
Preliminary upper limits on the North Fermi Bubble with HAWC-111 
data \cite{fermibubbleproceedings}.
The Bubble flux from \cite{fermibubblesfermicollab} is shown along with two extrapolations,
assuming it continues at E$^{-2}$ or E$^{-2.75}$. Preliminary
HAWC upper limits are
shown in black and rule out a pure E$^{-2}$ continuation of the Fermi 
Bubbles.
}
\label{fermibubbles}
\end{figure}

Finally, HAWC has a unique capability to search for large, extended TeV
emission from the Fermi Bubbles \cite{fermibubbles}, 
two large gamma-ray structures
extending above the center of the Galaxy. HAWC-111 data has been searched
and upper limits have been placed \cite{fermibubbleproceedings}.
Figure \ref{fermibubbles} shows the preliminary limits on the North Fermi
Bubble. HAWC limits rule out a pure E$^{-2}$ continuation of the spectrum.

\section{Extra-Galactic Science}
\label{extragalacticsection}

TeV photons produced outside our Galaxy may be attenuated by
pair production on
the Extra-Galactic Background Light (EBL) that fills the space between
galaxies. HAWC, with sensitivity to photons above 100 GeV,
maintains sensitivity to Extra-Galactic objects that
are nearby and bright, most notably Active Galactic Nuclei (AGN) 
and Gamma-Ray Bursts (GRBs). With HAWC, we monitor the sky for transient
emission from these objects. This monitoring is being conducted in real-time
at the HAWC site \cite{onlineblazarproceedings,grbuntriggeredproceedings}.

\begin{figure}
\includegraphics[width=0.95\linewidth]{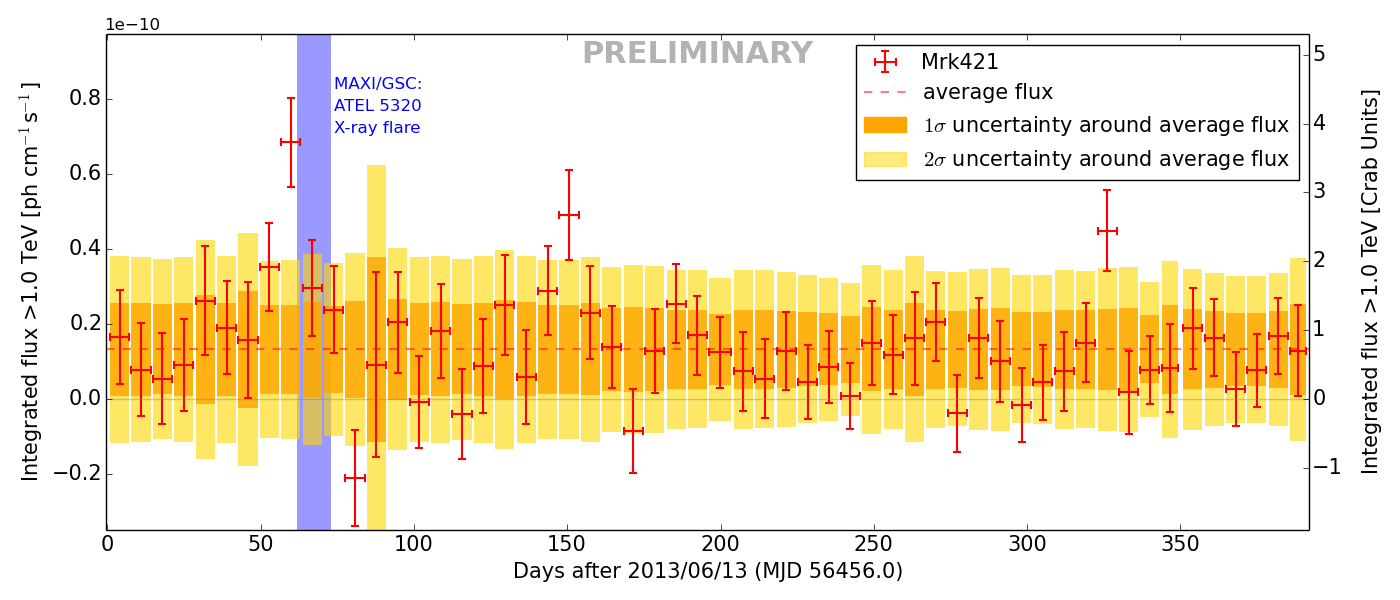}
\caption{
Flux light curve of Markarian 421 for the time between June 13, 2013, 
and July 9, 2014, in
intervals of 7 days, with horizontal error bars indicating the time 
span between the start of the first and the
end of the last transit over HAWC \cite{blazarproceedings}. 
Integrated fluxes above 1 TeV for a 
differential spectrum with power law
index 2.2 and exponential cut-off at 5 TeV are shown in photons per 
cm$^2$ per second on the left axis and divided by
the average Crab flux observed by HAWC on the right axis. 
The statistical 1 and 2 $\sigma$ flux uncertainties of
the individual measurements are displayed as orange/yellow bars around the 
weighted average flux (dashed
line). The large uncertainties in the flux around day $\sim$90 are due to a 
period of construction and maintenance
in September 2013 during which HAWC was shut down during day time for 
several days in a row, creating
large gaps in the transit coverage. The blue band indicates the period over 
which a strong increase in X-ray
flux was observed by MAXI \cite{maxiflare}.
}
\label{mrk421}
\end{figure}

Archival searches are also underway. In archival HAWC data, the 
blazars Markarian 421 and 501 show substantial, significant flux 
variability \cite{blazarproceedings}. Figure \ref{mrk421} shows an example from
this analysis, the long-term light curve of Markarian 421, in data from
HAWC-111, binned in week-long time bins.  
During the HAWC observation, the
blazar showed significant, substantial flux variability. Analysis is underway.

In addition to AGN, HAWC has the sensitivity to observe a nearby bright
Gamma-Ray Burst if
it is in the field-of-view of the instrument. Fermi observations show a
number of bursts (e.g. \cite{grb090510})
with a high-energy spectral component which is
rising up to the high-energy limits of Fermi, where the HAWC 
sensitive energy begins.
One estimate \cite{taboadaandgilmore},
extrapolating
GRB populations observed in Fermi, suggests HAWC might optimistically
observe 1-2
GRBs per year.

\begin{figure}
\includegraphics[width=0.95\linewidth]{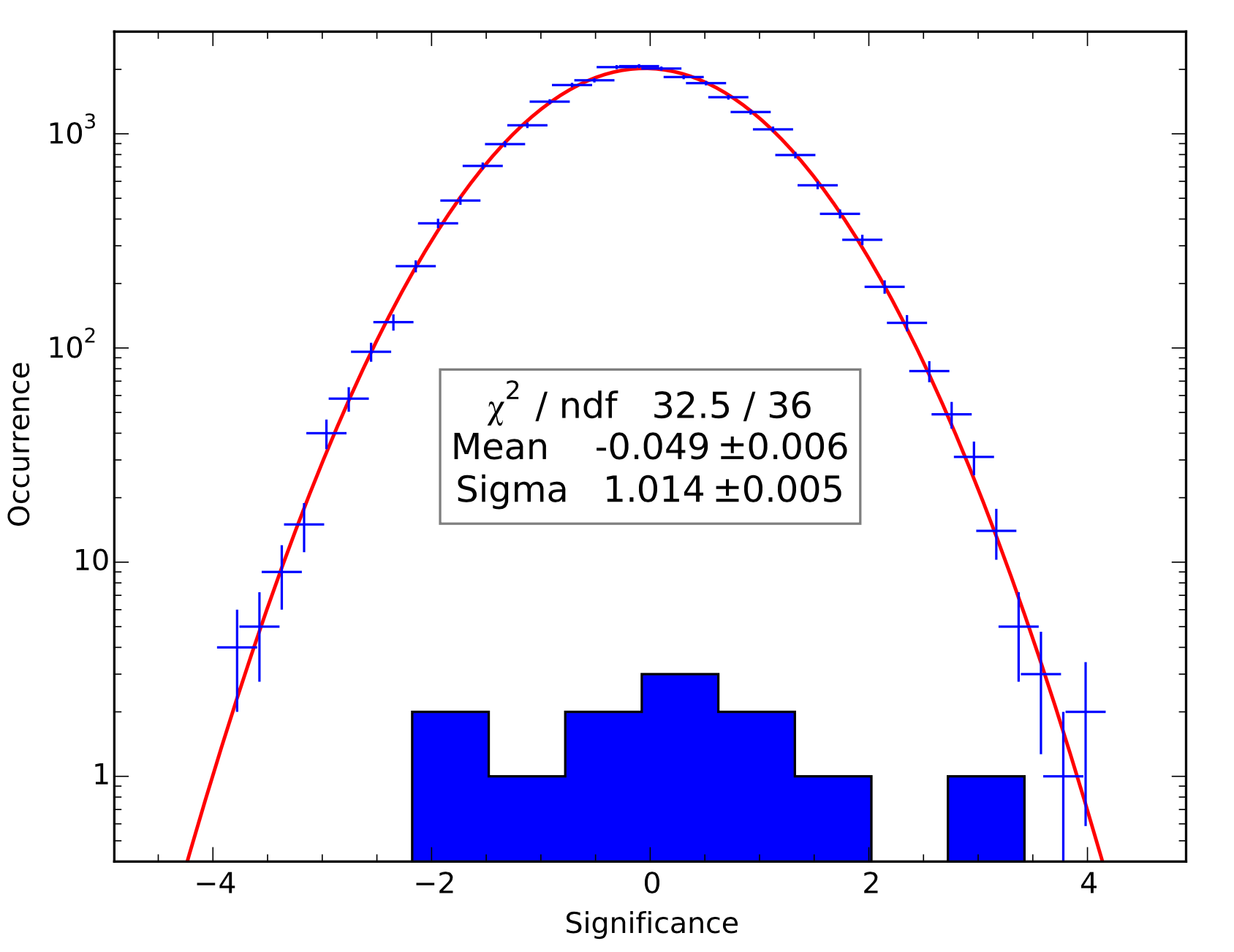}
\caption{
The significance distribution of 18 Swift-triggered GRBs as seen in 
HAWC (shaded blue) \cite{grbproceedings}. 
All events are consistent with pure background.
Additionally, the significance distribution of a large 
number of injected triggers is also shown, validating the computation
of the statistical significance. 
}
\label{grbs}
\end{figure}

During HAWC-111, 22 Swift-detected GRBs occurred in the HAWC field-of-view,
and we have data on 18 of them.
HAWC data was searched \cite{grbproceedings} 
for evidence of temporally coincident bursts of
events.
Observations were consistent with background. Figure \ref{grbs}
shows the significance distribution of the 18 GRBs in HAWC data along
with
random ``fake'' triggers, distributed as expected, to validate the
null hypothesis determination. 

\section {Fundamental Physics}
\label{fundamentalsection}

Understanding the particle nature of dark matter is one of the most 
pervasive questions in physics. If dark matter consists of multi-TeV
particles that can annihilate (predicted by many WIMP models), we expect
some level of gamma-ray production in dense regions of dark matter:
the Galactic Center, dark matter satellites of our galaxy, or nearby
galaxies or galaxy clusters. Because HAWC has a wide field-of-view, we can
search every candidate in our field-of-view. 

\begin{figure}
\includegraphics[width=0.95\linewidth]{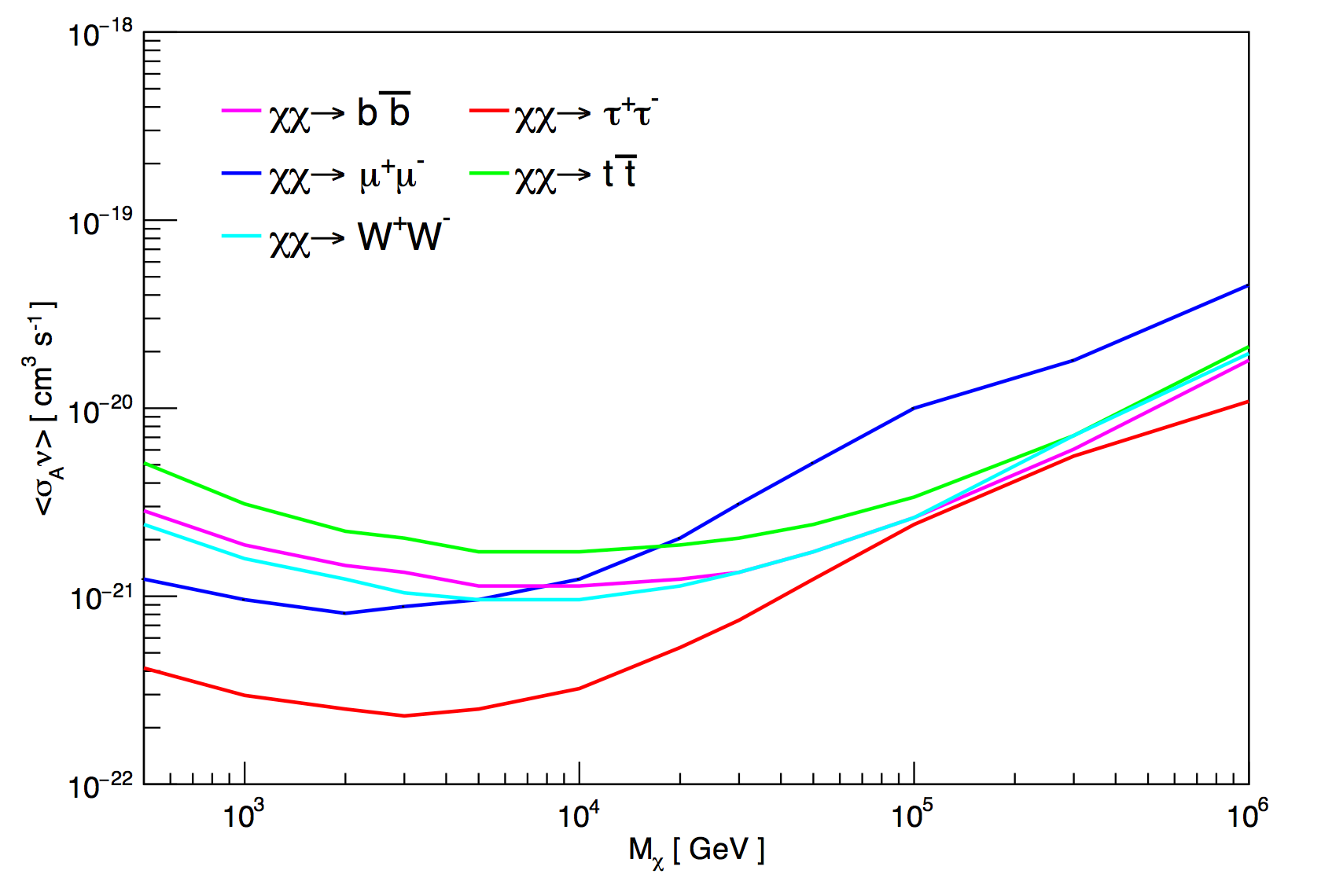}
\caption{
Combined annihilation cross-section limits from HAWC's non-observation of
photons of 14 dwarf spheroidal galaxies \cite{dmlimitsproceedings}. 
The velocity-weighted cross section limit is shown versus the mass of the dark
matter particle mass. The limit depends on the assumed annihilation 
products.
}
\label{dmlimits}
\end{figure}

Figure
\ref{dmlimits} shows the results of a search for a dark matter signal
from dwarf galaxies in the field-of-view of HAWC using 180 days of
the HAWC-111 dataset.
The dwarf galaxies have been individually searched for a photon excess,
and combined into a global upper limit
\cite{dmlimitsproceedings}.
Eventually HAWC limits will be competitive at extremely high masses above 
10-1000 TeV \cite{dmproceedings}. HAWC will also be particularly sensitive to decaying dark matter
and dark matter in galaxy clusters, where the sources are expected to be 
more extended.

HAWC is uniquely suited to search for bursts of TeV photons
from primordial black hole (PBH) evaporation \cite{pbhproceedings,milpbh}. 
The photon emission from PBHs is broad-band \cite{pbhmodel} 
and may be difficult 
to distinguish from a Gamma-Ray Burst. If a PBH evaporation event does
occur, they must be in the vicinity of the Earth where the EBL attenuation of
TeV photons will not be present, as it is for GRBs. 
As such, if HAWC sees multi-TeV photons
of the correct spectral and temporal shape, we can positively identify
an evaporation event.


A suitable GRB, at sufficiently high redshift and with a sufficiently hard
spectrum, if observed by HAWC, could be used to derive limits on 
Lorentz Invariance Violation:
energy-dependent variations of the speed of light. HAWC's higher reach in 
energy, compared to satellite-based instruments, provides it with the ability
to set competitive limits \cite{livproceedings}. 


\section {Cosmic-Ray Science}
\label{crsection}

Finally, HAWC serves as a unique observatory for studying cosmic rays 
that arrive at Earth, studying what is normally regarded as our background
to gamma-ray observation.
In particular, HAWC triggers at $\sim$15kHz and will collect more 
than a trillion cosmic rays between 100 GeV and 100 TeV before the experiment
is finished. 
In one second, we resolve the relative cosmic-ray flux to about
1\%, and minute fluctuations in the cosmic-ray flux, either due to
solar modulation \cite{solarproceedings} or due to spatial fluctuations
in the cosmic-ray background \cite{cranisotropyproceedings}
can be seen.

\begin{figure}
\includegraphics[width=0.95\linewidth]{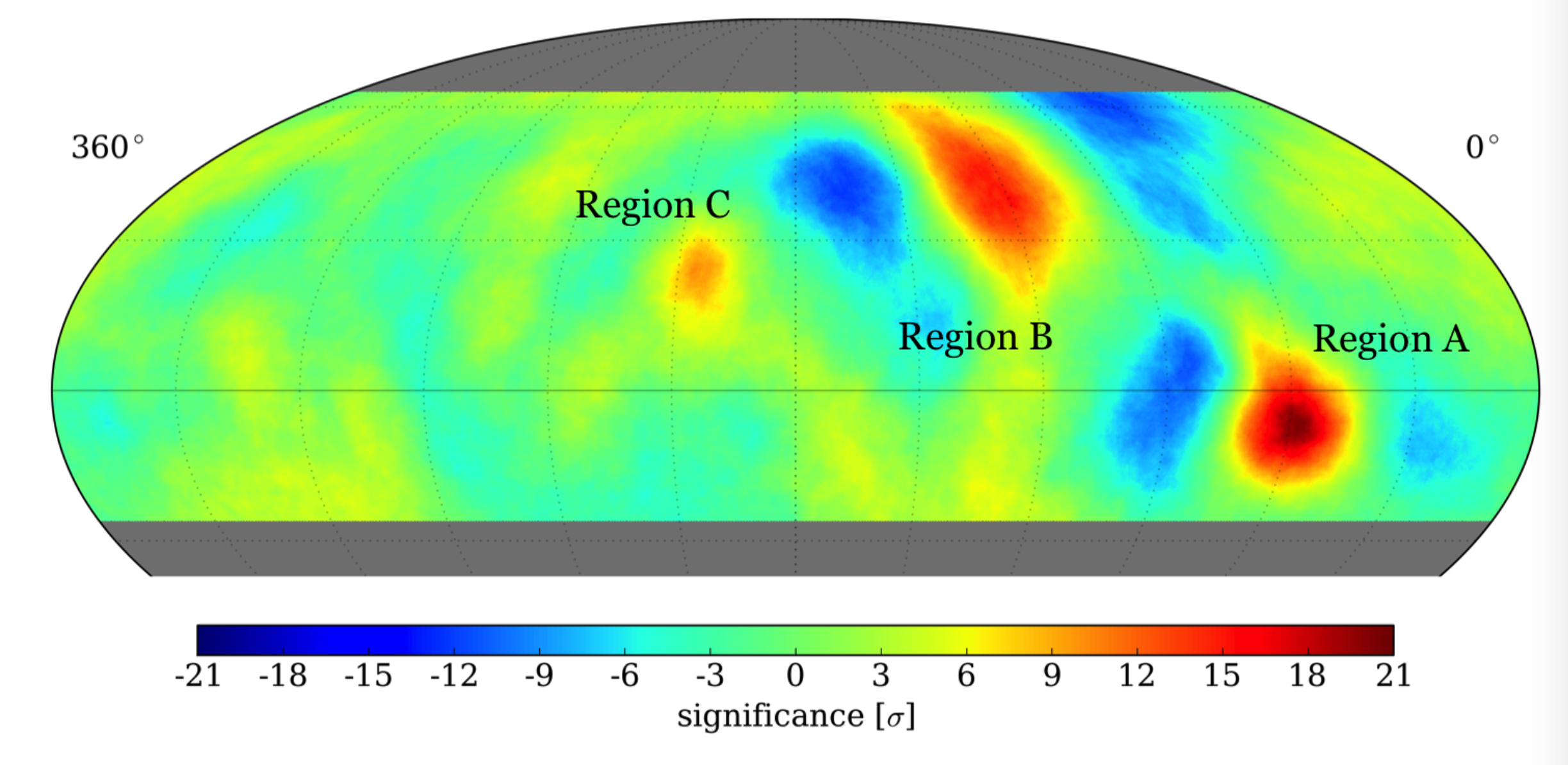}
\caption{
The HAWC cosmic-ray skymap, using data from HAWC-111, 
in equatorial coordinates \cite{cranisotropyproceedings}. 
The map has had large-scale dipole, quadrupole
and octopole components subtracted, and a 10$^\circ$ smoothing applied.
Milagro's Regions A and B are seen, along with a new Region C.
}
\label{cranisotropy}
\end{figure}

Figure \ref{cranisotropy} shows the HAWC view of the 10$^\circ$-scale 
cosmic-ray anisotropy first discovered in Milagro data 
\cite{milagrolocalizedexcess}. These anisotropies are small,
a few parts in $10^4$. HAWC data confirms the Milagro detections: 
``Region A'' and
``Region B'', and adds to it a third detected excess termed ``Region C''
first reported in ARGO-YBJ data \cite{cranisotropyargo}.
With increasing sensitivity, it is becoming possible to investigate
the spectral properties of the anisotropy \cite{cranisotropyproceedings}
and to cross-correlate with anisotropy seen by IceCube cosmic-ray
observations in the south \cite{allskycranisotropyproceedings}.

HAWC is such a precise cosmic-ray barometer that it is possible to 
study the local cosmic-ray modulation by the sun \cite{solarproceedings}
and to image the cosmic-ray silhouette of the sun \cite{sunshadowproceedings},
allowing a time-dependent picture of the magnetic field of the sun.

Finally, HAWC's photon/hadron separation admits the possibility to study
cosmic-ray electrons \cite{segevsproceedings}. 
Cosmic-ray electrons will induce pure electro-magnetic showers, just
like photons, but will appear as a large diffuse population, not localized
like photons. 

\section{Summary and Outlook}

HAWC brings new capabilities to TeV astronomy and astrophysics. With its
wide field-of-view and continuous operation, HAWC is poised to discover
new sources and continuously monitor the sky for rare transient events
across the Northern sky. Interestingly enough, the Southern sky has never
been surveyed with a wide-field TeV instrument. Such an
instrument \cite{southproceedings} is under consideration.
With HAWC we will continuously survey the sky
for the highest-energy photons ever detected and are poised to revolutionize
the field of high-energy astrophysics.

\section*{Acknowledgments}
\footnotesize{
We acknowledge the support from: the US National Science Foundation (NSF);
the US Department of Energy Office of High-Energy Physics;
the Laboratory Directed Research and Development (LDRD) program of
Los Alamos National Laboratory; Consejo Nacional de Ciencia y Tecnolog\'{\i}a (CONACyT),
Mexico (grants 260378, 55155, 105666, 122331, 132197, 167281, 167733);
Red de F\'{\i}sica de Altas Energ\'{\i}as, Mexico;
DGAPA-UNAM (grants IG100414-3, IN108713,  IN121309, IN115409, IN111315);
VIEP-BUAP (grant 161-EXC-2011);
the University of Wisconsin Alumni Research Foundation;
the Institute of Geophysics, Planetary Physics, and Signatures at Los Alamos National Laboratory;
the Luc Binette Foundation UNAM Postdoctoral Fellowship program.
}

\bibliographystyle{JHEP}
\bibstyle{JHEP}

\bibliography{bibliography}

\end{document}